\documentclass{ecai}

\usepackage{graphicx}
\usepackage{latexsym}

\usepackage[super]{nth}
\usepackage{float} 

\begin{document}

\begin{frontmatter}

\title{Analysis of Knowledge Tracing performance on synthesised student data}

\author[A]{\fnms{Panagiotis}~\snm{Pagonis}\thanks{Corresponding Author. Email: Panos.Pagonis@thi.de.}}
\author[A]{\fnms{Kai}~\snm{Hartung}}
\author[A]{\fnms{Di}~\snm{Wu}} 
\author[A,B]{\fnms{Munir}~\snm{Georges}}
\author[A]{\fnms{Sören}~\snm{Gröttrup}}

\address[A]{Technische Hochschule Ingolstadt, Germany}
\address[B]{Intel Labs, Germany}

\begin{abstract}

Knowledge Tracing (KT) aims to predict the future performance of students by tracking the development of their knowledge states. Despite all the recent progress made in this field, the application of KT models in education systems is still restricted from the data perspectives: 1) limited access to real life data due to data protection concerns, 2) lack of diversity in public datasets, 3) noises in benchmark datasets such as duplicate records. To resolve these problems, we simulated student data with three statistical strategies based on public datasets and tested their performance on two KT baselines. 
While we observe only minor performance improvement with additional synthetic data, our work shows that using only synthetic data for training can lead to similar performance as real data.
\end{abstract}

\end{frontmatter}

\section{Introduction}

In recent years, the education has experienced the transformation from traditional classrooms into digital teaching \cite{8924064}. Especially the COVID-19 pandemic remarkably accelerated the adaption of remote learning and encouraged the development of AI-enhanced educational technology. Intelligent tutoring systems (ITS) are widely applied in providing personalized instruction and tutoring for students with computer-assisted learning environments, which offer intelligent services that assist the knowledge acquisition process and increase motivation. However, online education still faces the challenges of improving learner engagement and adapting educational contents to different users. As a sequence modeling technique, Knowledge Tracing (KT) models dynamically keep track of students' knowledge states and predict their future performance on following exercises. This has been proven to be valuable in supporting instructors with better teaching strategy design and creating personalised study plans in accordance with learner's diverse learning styles \cite{liu2021survey}.

As a student modeling technique, the KT task aims to observe and capture the changes in students' knowledge states during learning. Given series of questions, the learning process is represented through the results, corresponding skills and interaction data, like response time, attempts, or skills, of each question answering activity between the ITS and the learners. 
A large amount of efforts has been made to study this problem. Bayesian Knowledge Tracing (BKT) \cite{corbett1994knowledge} is the most classic approach, which applies Hidden Markov Models (HMM) to predict probabilities of the binary variable of each step. In its earlier versions, this method assumes that studied knowledge will never be forgotten and ignores the individual differences of students' prerequisite knowledge, and therefore has led to the emergence of multiple varieties, see \cite{kaser2017dynamic, yudelson2013individualized}. With the rise of neural networks, Deep Knowledge Tracing (DKT) \cite{piech_DKT_2015} introduced recurrent neural networks as a stronger sequential feature extractor, which outperformed previous approaches. New DKT models began to be proposed to capture different aspects of the cognitive development process, see, for example, \cite{ghosh2020context,liu2019ekt,nakagawa2019graph}.

Though, a lot of efforts have been made to improve the prediction accuracy on student grades, with most KT models focusing on binary grade prediction and only care if a student passed or failed an exercise, see \cite{badrinath2021pybkt, dai2021knowledge, lee_NPAKT_2019, liu_QEmbKT_2020, pandey_SAKT_2019, piech_DKT_2015}.
Additionally, most common public datasets like Assistments \cite{feng_assistment_2009, pardos_assistment_2014}, Junyi \cite{Chang_JunYi_2015} or EdNet \cite{choi_ednet_2020} also provide data with only binary grades of exercises.
In practical application, however, exercises with grades on some kind of scale beyond just 0 and 1 are  quite common. So, the lack of more diverse data sources is significant in this area. Furthermore, currently available public KT datasets are collected from diversed learning scenarios, which present distinctively different features and scales, resulting in a difficulty of practical adaption. On one hand, the benchmark datasets often contain noise (e.g. wrong timestamps or null values), formatting issues (e.g. gaps between experiment setups, datasets, and versions), and incorrect descriptions.
Furthermore, strict data protection regulations often  challenge the researchers during the data collection. Despite of the trending of digital learning, the use of learning analytic tools are strictly limited and many developers, like in the European Union, do not have access to anonymized student data \cite{GDPR2016a}. It is worth noting that the majority work focuses on modeling the learning process instead of solving the mentioned problems from a data perspective. As a result, conducting knowledge tracing in real education could be very expensive.

Seeking solutions for the lack of variety and expensive data in KT, we focus our efforts in synthesizing data on continuous or ordinal scales and test their performance on two baselines (see Section 4) by adding synthesized data of different ratio to the original datasets.
We accordingly chose two public datasets for our experiments which do record grades on such scales, the Open University Learning Analytics dataset (OULAD) \cite{kuzilek_2017_oulad},  and the SLP dataset \cite{yu_2021_slp}.

\begin{table}[t!h]
    \centering
    \begin{tabular}{l|cc}
         & \textbf{OULAD} & \textbf{SLP} \\
        \hline
        students & 32,593 & 4,830 \\ 
        exercises & 37,443 & 7,502 \\ 
        total interactions & 10,655,280 & 460,688 \\ 
        \hline
    \end{tabular}
    \caption{Size statistics of the OULAD \cite{kuzilek_2017_oulad} and SLP \cite{yu_2021_slp} datasets}
    \label{table:datasets}
\end{table}

Both datasets contain interaction data from online learning environments.
The size description can be found in Table \ref{table:datasets}.
OULAD gives grades ranging from 0 to 100 for each exercise and SLP gives ordinal grades with the range depending on each exercise.
To be able to treat the datasets equally, we normalize all grades in terms of percentage to a range from 0 to 100.

In the following section, we describe our approaches to generate synthetic data. 
In Section \ref{sec:KT} and \ref{sec:exp}, we describe the KT baselines and the experiments using the corresponding models on the synthetic data.
Section \ref{sec:disc} and \ref{sec:conclusion} are dedicated to the discussion and conclusion of our findings.

\section{Synthetic Data Generator}

The most crucial parameter in the educational datasets is the students' grades. We designed three different methods to generate grades, which are introduced in this section: Generator$_1$ is based on the distribution of grades, Generator$_2$ performs resampling in each step of the student learning path, the sequence of grades from one student, and Generator$_3$ reuses the existing students. When examining the datasets for the generator creation, we exclude grades 0 and 100 to avoid a dominant influence from those two values.

\subsection{Generator$_1$}
Synthetic Data Generator$_1$ generates new grades by sampling from one distribution of all grades in the dataset. So, we first identify the best fitting distribution to the grades in the real data. 
We did so using the 'distfit' package in python \cite{taskesen_erdogan_2023_7767074}, which comprises nearly all common univariate distributions.

\begin{figure}[h]
    \centering
    \includegraphics[scale=0.24]{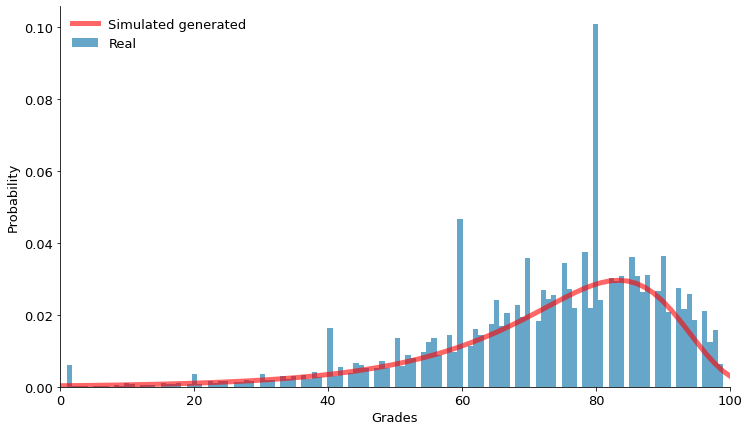}
    \caption{Fitted log-gamma distribution of grades (curve) compared to distribution of all grades (histogram) for the OULAD dataset.}
    \label{fig:stats_fig1}
\end{figure}

Specifically, in the OULAD dataset the best fitting distribution to the grades is a log-gamma distribution with parameters: $c=0.50,$ $scale=8.20,$ $location=89.02$, see Figure \ref{fig:stats_fig1}.

By drawing grades from this distribution, we create the single interaction instances for the synthetic dataset.

\begin{figure}[h]
    \centering
    \includegraphics[scale=0.24]{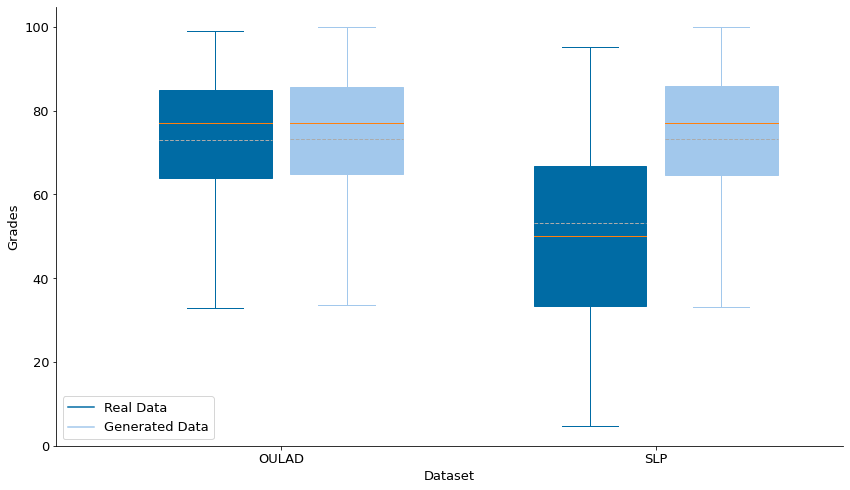}
    \caption{Boxplot of real and synthetic grade distribution using Generator$_1$ for OULAD and SLP dataset.}
    \label{fig:stats_fig3}
\end{figure}

The real and the simulated grades datasets have a similar distributions, as visualized in Figure \ref{fig:stats_fig3}. Additionally, both datasets have a similar average (real: $73.02$, simulated: $73.14$) and median (real: $77$, simulated: $76.98$) grade, as well as standard deviation (real: $17.49$, simulated: $17.67$). 

\begin{figure}[h]
    \centering
    \includegraphics[scale=0.24]{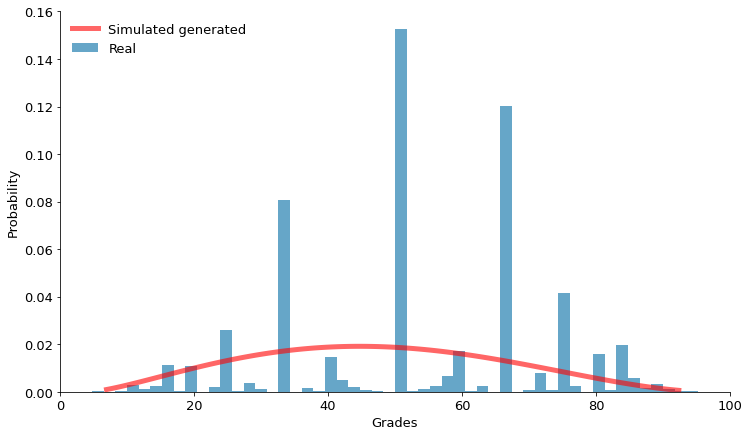}
    \caption{Fitted beta distribution of grades (curve) compared to the distribution of all grades (histogram) for the SLP dataset.}
    \label{fig:stats_fig2}
\end{figure}

In the SLP dataset, the distribution that explains the grades best is a beta distribution with parameters: $a=2.46,$ $b=2.89,$ $scale=92.62,$ $location=4.38$. As shown in Figure \ref{fig:stats_fig2}, the selected distribution is not optimal.
It should be mentioned that there is not always a perfect fit of distributions.

The average (real: $53.16$, simulated: $73.17$) and the median (real: $50.00$, simulated: $77.03$) are different, but the standard deviation (real: $18.48$, simulated: $17.17$) is still similar, which is shown in Figure \ref{fig:stats_fig3}.

Apart from the basic descriptive statistics one more parameter that should be taken into account, is the possible dependency on the previous grade. In Generator$_1$ this dependency is not taken into account.
The development of a function that takes into account the current attempt of a student and the distribution of grades at that point should give very reliable results. However, this needs a lot of data which radically increases the complexity. Generator$_1$ can be considered as the baseline for our next attempts to provide a powerful and reliable method for synthetic datasets. 

\subsection{Generator$_2$}
Another approach to generate synthetic educational data is to perform a resampling from the actual dataset. 
This idea is inspired by bootstrap resampling for estimating parameters of a dataset \cite{efron1983estimating,efron1997improvements}.
Usually, bootstrapping is used to estimate the variance or, in general, the distribution of a statistic of a population such as the mean or variance.
However, directly applying this concept would not produce realistic grades.
Thus, we don't do any actual parameter estimation, but only use the approach to collect samples.

\begin{figure}[h]
    \centering
    \includegraphics[scale=0.24]{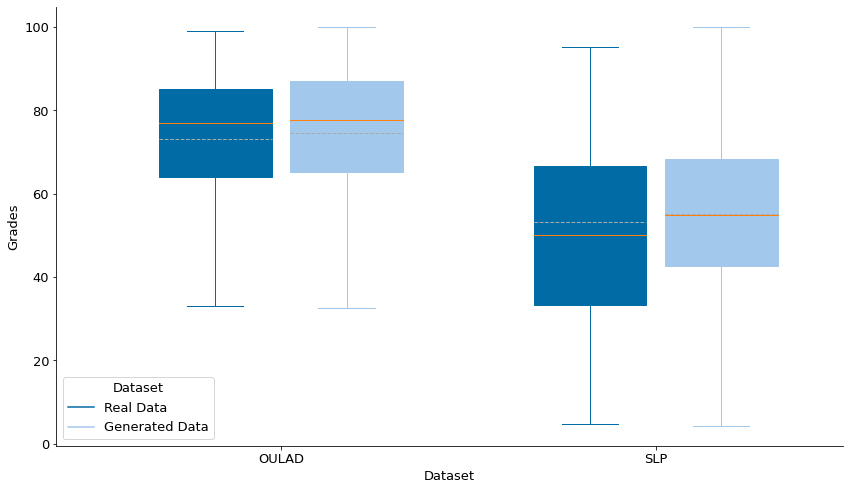}
    \caption{Boxplot of real and synthetic grade distribution using Generator$_2$ for OULAD and SLP dataset.}
    \label{fig:stats_fig4}
\end{figure}

First, we randomly collect the whole learning path of a user, which is the sequence of grades this student got during its study.
Then, for each step in this learning path, we randomly pick a value out of the grades over all students at the same step from the real data.
Thus, this method does not cover progressive dependencies between the subsequent time steps.
Afterwards we add some Gaussian noise, with parameters $\mu=0, \sigma=3$, to create the new grade. This process is repeated until all grades in every step of the learning path are filled. This new synthetic dataset has similar properties and a similar grade distribution to the real one. The advantage of this method is the construction of the dataset based on real data and considering each step.
Using as input for the Generator$_2$ the OULAD dataset, the results are encouraging. A comparison to the descriptive statistics of real and generated data proves that efficiency. More specifically, the average (real: $73.02$, simulated: $74.43$), the median (real: $77.00$, simulated: $77.74$) and the standard deviation (real: $17.49$, simulated: $16.81$) do not differ much. These comparisons is visualized in Figure \ref{fig:stats_fig4}.

Moreover, in the SLP dataset the similarities in the main metrics average (real: $53.16$, simulated: $55.11$), median (real: $50.00$, simulated: $54.79$) and standard deviation (real: $18.48$, simulated: $18.05$) remain, but the first quartile (real: $33.33$, simulated: $42.61$) differs slightly. Those results can be found in Figure \ref{fig:stats_fig4}.

There are no limits in the dataset size, thus any source with numeric grades may work without issues. A possible drawback of this method is the time of generation, which can be high depending on the requested amount of data.

\subsection{Generator$_3$}

The third version of the Synthetic Data Generator or Generator$_3$ utilises a simpler and the more lightweight approach. This method chooses a random user from the real dataset and collects their whole learning path. Then, Gaussian noise is added to the grades, here with parameters $\mu=0, \sigma=3$. The procedure continues until the requested amount of data is fulfilled. 
The method offers a quick generation of any amount of data which preserves the properties of the real data, as well as the possible time dependency between grades and correlations. Those are not the only advantages of the Generator$_3$, but also it could be easily modified and used for any dataset. Even if the data are reused, it offers anonymity and generated users could be considered as completely new entries.

\begin{figure}[h]
    \centering
    \includegraphics[scale=0.24]{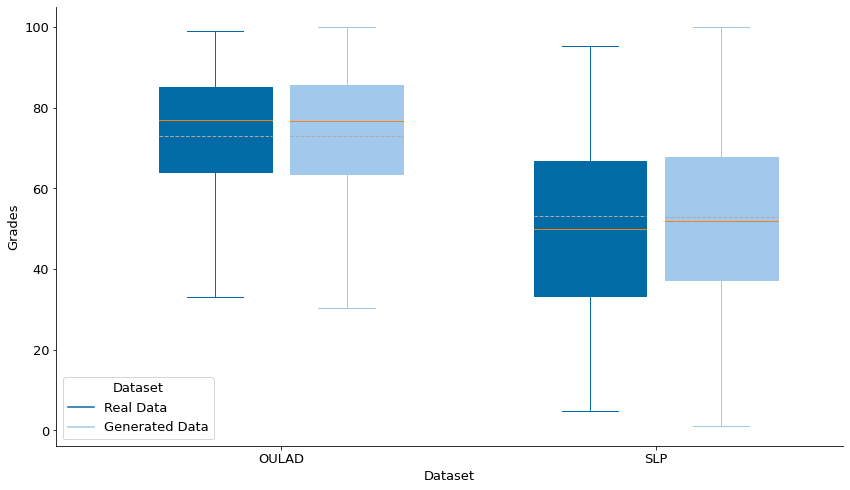}
    \caption{Boxplot of real and synthetic grade distribution using Generator$_3$ for OULAD and SLP dataset.}
    \label{fig:stats_fig5}
\end{figure}

For the evaluation of the algorithm the same two datasets are used. A comparison between real and simulated data follows for both datasets as well as some basic statistics.
In the OULAD dataset the descriptive statistics of real and generated datasets are almost identical. The difference for the average (real: $73.01$, simulated: $73.03$), the median (real: $77.00$, simulated: $77.68$) and the standard deviation (real: $17.48$, simulated: $17.70$) are almost equal to 0 as can be seen in Figure \ref{fig:stats_fig5}.
The method has similar results on the SPL dataset, which are also shown in Figure \ref{fig:stats_fig5}. Here, the average (real: $76.89$, simulated: $73.03$), the median (real: $50.00$, simulated: $51.96$) and the standard deviation (real: $18.47$, simulated: $18.70$) are once again very similar.
In particular, the OULAD generated dataset follows a beta distribution with parameters $a=25.70$, $b=2.37$, $scale=343.82$, $location=-241.75$, which is very close to the log-gamma observed for the original dataset, but more smoothed out.
The SLP follows a beta distribution with parameters $a=3.37$, $b=2.98$, $scale=100.73$, $location=-0.68$.

\begin{figure}[h]
    \centering
    \includegraphics[scale=0.28]{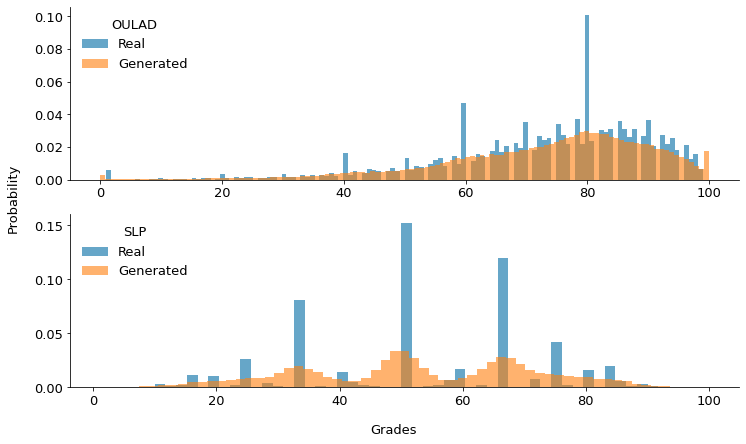}
    \caption{Comparison of the distributions of grades of real data and data generated from Generator$_3$.}
    \label{fig:stats_fig6}
\end{figure}

In addition, one more proof for the efficiency of that method is that the distributions of the generated data follow similar but smoother distributions as the real data (see Figure \ref{fig:stats_fig6}).

\section{Knowledge Tracing} \label{sec:KT}

We compare two approaches to Knowledge Tracing.
The original Deep Knowledge Tracing (DKT) \cite{piech_DKT_2015} uses an recurrent neural network architecture, which takes a sequence of binary grades of one student as input and gives a prediction on whether the next exercise will be passed or failed as output.
We adapt this architecture to the non-binary grades of our datasets.
Instead of cross-entropy loss we apply mean-squared-error as loss, comparing the networks logits scaled by a sigmoid function with the grades normalized to the range $[0,1]$.

Bayesian Knowledge Tracing (BKT), as implemented by Badrinath et al. in \cite{badrinath2021pybkt}, uses hidden Markov models  optimized via expectation maximization.
We are running the default BKT model as offered by the library with the additon of the forget parameter.

\section{Experiments} \label{sec:exp}

For both datasets, we compare the influence of adding synthetic data to real training data as augmentation, as well as the performance of the Knowledge Tracing model on real data when trained on synthetic data only.
The test data is the same 20\% split of the full real data for all experiments on one dataset.

Of the 80\% training data split, we compare KT performance when training on 0\%, 50\%, and 100\% of that split respectively.
For each of these variants we further compare performance when adding synthesized data of a size equal to 0\%, 25\%, 50\%, 75\%, 100\%, 200\% and 300\% of the original training split.
Thus we get a grid of real and synthetic data ratios on which to train the KT models and for which to compare the result.
%We do exclude the variant with 0\% real and synthetic training data, as that is just no training at all.
This results in two variations for each dataset without synthetic data, and 18 more variations for each dataset and generator version.

\subsection{Evaluation}

We evaluate the KT models on three metrics, Accuracy (ACC), Matthews-Correlation-Coefficient\cite{matthews_1975} (MCC) and Mean Absolute Error (MAE).
MAE serves to asses the Model Performance with respect to the continuous nature of the scores in the datasets we chose to focus on.
ACC evaluates the classic binary prediction of passed or failed and MCC should give an impression on how balanced the model predictions are.
For both classification scores, we treat a normalized score above 0.5 as class 1 (passed) and a score below as class 0 (failed).
We don't use the ordinal grades of the SLP dataset as categories as they do not provide consistent classes across the dataset.

\subsection{Results}

\begin{figure*}[ht!]
    \centering
    \includegraphics[scale=0.58]{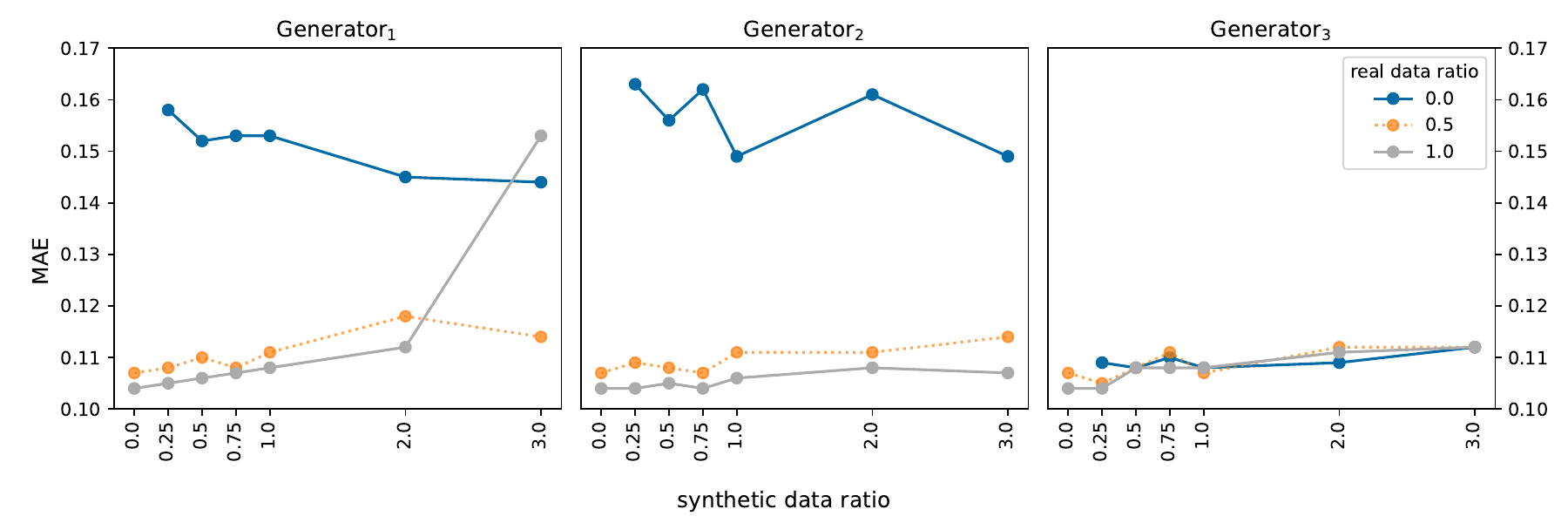}
    \caption{OULAD Dataset: Influence of synthetic data on Deep Knowledge Tracing performance in MAE.
    Lower values are better.}
    \label{fig:oulad_mae}
\end{figure*}

\begin{figure*}[ht!]
    \centering
    \includegraphics[scale=0.58]{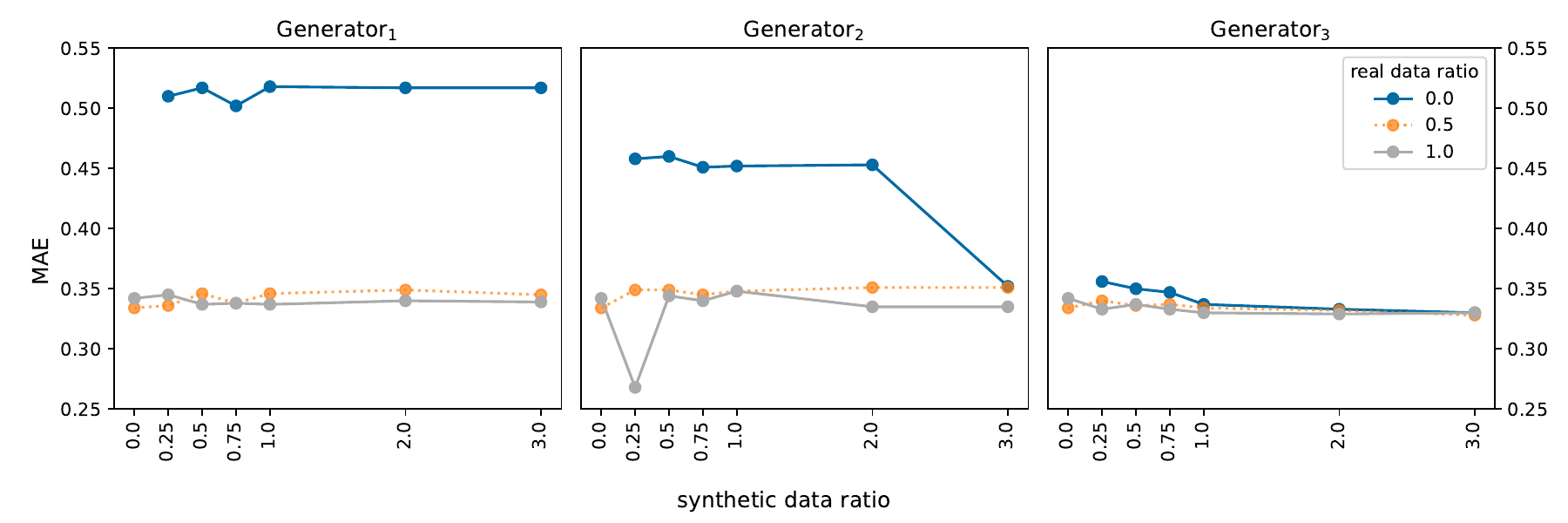}
    \caption{SLP Dataset: Influence of synthetic data on Deep Knowledge Tracing performance in MAE.
    Lower values are better.}
    \label{fig:SLP_mae}
\end{figure*}

Generator$_1$ mainly worsens the model performance of DKT on the OULAD dataset to a small degree when added to real training data.
When using only synthetic data, the MAE is worse than when training on any amount of real data, however only by 3-4 points, see left graph in Figure \ref{fig:oulad_mae}.
On the other hand, when looking at the MCC in Table \ref{table:mcc_oulad}, the synthetic data stays well below the real data.
On the SLP dataset adding synthetic data from Generator$_1$ has less of a negative effect, but when using synthetic data only the MAE is further away from the baseline of real data than on OULAD, see Figure \ref{fig:SLP_mae} and Table \ref{table:mcc_SLP}.

\begin{figure*}[ht!]
    \centering
    \includegraphics[scale=0.58]{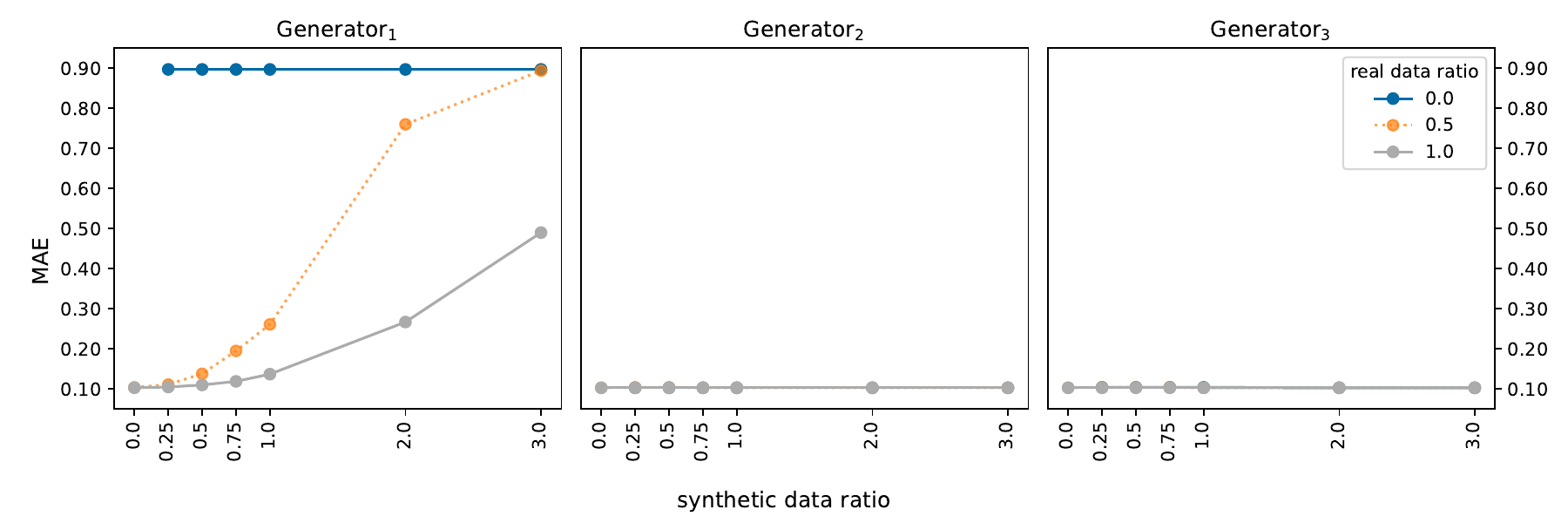}
    \caption{SLP Dataset: Influence of synthetic data on Bayesian Knowledge Tracing performance in MAE.
    Lower values are better.}
    \label{fig:SLP_mae_BKT}
\end{figure*}

\begin{table}[!th]
    \centering
    \begin{tabular}{l||c|cccccc}
        & \textbf{0} & \textbf{0.25} & \textbf{0.5} & \textbf{0.75} & \textbf{1} & \textbf{2} & \textbf{3}\\
        \hline\hline

        \multicolumn{8}{c}{Generator$_1$} \\
        \hline
     
        \textbf{0.0} & & 0.906 & 0.91 & 0.91 & 0.91 & \textbf{0.911} & \textbf{0.911} \\
        \textbf{0.5} & \textbf{0.92} & 0.918 & 0.919 & 0.918 & 0.917 & 0.917 & 0.918 \\
        \textbf{1.0} & 0.919 & \textbf{0.921} & 0.919 & 0.919 & 0.917 & 0.918 & 0.91 \\
        \hline \hline
        \multicolumn{8}{c}{Generator$_2$} \\
        \hline
     
        \textbf{0.0} &  & 0.898 & 0.909 & 0.892 & \textbf{0.911} & 0.896 & \textbf{0.911} \\
        \textbf{0.5} & \textbf{0.92} & \textbf{0.92} & 0.919 & 0.918 & 0.918 & 0.918 & 0.918 \\
        \textbf{1.0} & 0.919 & \textbf{0.921} & \textbf{0.921} & 0.92 & 0.919 & 0.92 & 0.919 \\
        \hline \hline
        \multicolumn{8}{c}{Generator$_3$} \\
        \hline
     
        \textbf{0.0} & & 0.912 & \textbf{0.915} & 0.91 & \textbf{0.915} & 0.913 & 0.909 \\
        \textbf{0.5} & \textbf{0.92} & 0.919 & 0.914 & 0.911 & 0.915 & 0.908 & 0.907 \\
        \textbf{1.0} & 0.919 & \textbf{0.92} & 0.914 & 0.914 & 0.915 & 0.911 & 0.909 \\
    \end{tabular}
    \caption{DKT \textbf{ACC} per data combination on the \textbf{OULAD} dataset. 
    Rows stand for real data ratio, columns for synthetic data ratio. 
    Best values per real data ratio are highlighted.}
    \label{table:acc_oulad}
\end{table}

\begin{table}[!th]
    \centering
    \begin{tabular}{l||c|cccccc}
        & \textbf{0} & \textbf{0.25} & \textbf{0.5} & \textbf{0.75} & \textbf{1} & \textbf{2} & \textbf{3}\\
        \hline\hline
        \multicolumn{8}{c}{Generator$_1$} \\
        \hline
        \textbf{0.0} & & -0.003 & 0.0 & 0.0 & \textbf{0.0003} & 0.0 & 0.0 \\
        \textbf{0.5} & 0.356 & 0.339 & 0.336 & 0.307 & 0.326 & 0.265 & \textbf{0.36} \\
        \textbf{1.0} & \textbf{0.385} & 0.36 & 0.329 & 0.342 & 0.335 & 0.3 & 0.328 \\
        \hline \hline
        \multicolumn{8}{c}{Generator$_2$} \\
        \hline
        \textbf{0.0} & & \textbf{0.004} & -0.004 & -0.019 & 0.0 & -0.007 & 0.0 \\
        \textbf{0.5} & 0.356 & 0.349 & \textbf{0.359} & 0.354 & 0.323 & 0.324 & 0.313 \\
        \textbf{1.0} & \textbf{0.385} & 0.381 & 0.376 & 0.376 & 0.338 & 0.33 & 0.333 \\
        \hline \hline
        \multicolumn{8}{c}{Generator$_3$} \\
        \hline
        \textbf{0.0} & & 0.344 & 0.347 & 0.344 & 0.359 & \textbf{0.361} & 0.335 \\
        \textbf{0.5} & 0.356 & \textbf{0.374} & 0.367 & 0.35 & 0.367 & 0.329 & 0.337 \\
        \textbf{1.0} & 0.385 & \textbf{0.39} & 0.371 & 0.359 & 0.368 & 0.354 & 0.34 \\
    \end{tabular}
    \caption{DKT \textbf{MCC} per data combination on the \textbf{OULAD} dataset. 
    Rows stand for real data ratio, columns for synthetic data ratio. 
    Best values per real data ratio are highlighted.}
    \label{table:mcc_oulad}
\end{table}

\begin{table}[!th]
    \centering
    \begin{tabular}{l||c|cccccc}
        & \textbf{0} & \textbf{0.25} & \textbf{0.5} & \textbf{0.75} & \textbf{1} & \textbf{2} & \textbf{3}\\
        \hline\hline
        \multicolumn{8}{c}{Generator$_1$} \\
        \hline
        \textbf{0.0} & & 0.489 & 0.468 & \textbf{0.493} & 0.468 & 0.467 & 0.467 \\
        \textbf{0.5} & \textbf{0.706} & 0.69 & 0.695 & 0.693 & 0.689 & 0.692 & 0.692 \\
        \textbf{1.0} & 0.7 & \textbf{0.703} & 0.699 & 0.702 & 0.702 & 0.702 & \textbf{0.703} \\
        \hline \hline
        \multicolumn{8}{c}{Generator$_2$} \\
        \hline
        \textbf{0.0} & & 0.503 & 0.499 & 0.502 & 0.504 & 0.506 & \textbf{0.691} \\
        \textbf{0.5} & \textbf{0.706} & 0.682 & 0.665 & 0.669 & 0.698 & 0.674 & 0.675 \\
        \textbf{1.0} & 0.7 & \textbf{0.78} & 0.692 & 0.713 & 0.7 & 0.707 & 0.705 \\
        \hline \hline
        \multicolumn{8}{c}{Generator$_3$} \\
        \hline
        \textbf{0.0} & & 0.662 & 0.692 & 0.692 & \textbf{0.698} & 0.696 & 0.69 \\
        \textbf{0.5} & \textbf{0.706} & 0.702 & 0.698 & 0.697 & \textbf{0.706} & 0.695 & 0.695 \\
        \textbf{1.0} & 0.7 & \textbf{0.709} & 0.707 & 0.707 & 0.704 & 0.705 & 0.704 \\
    \end{tabular}
    \caption{DKT \textbf{ACC} per data combination on the \textbf{SLP} dataset. 
    Rows stand for real data ratio, columns for synthetic data ratio. 
    Best values per real data ratio are highlighted.}
    \label{table:acc_SLP}
\end{table}

\begin{table}[!th]
    \centering
    \begin{tabular}{l||c|cccccc}
        & \textbf{0} & \textbf{0.25} & \textbf{0.5} & \textbf{0.75} & \textbf{1} & \textbf{2} & \textbf{3}\\
        \hline\hline
        \multicolumn{8}{c}{Generator$_1$} \\
        \hline
        \textbf{0.0} &  & -0.04 & \textbf{0.04} & -0.05 & 0.017 & 0.004 & 0.001 \\
        \textbf{0.5} & \textbf{0.412} & 0.381 & 0.389 & 0.384 & 0.383 & 0.385 & 0.385 \\
        \textbf{1.0} & 0.397 & \textbf{0.409} & 0.403 & 0.404 & 0.403 & 0.408 & \textbf{0.409} \\
        \hline \hline
        \multicolumn{8}{c}{Generator$_2$} \\
        \hline
        \textbf{0.0} &  & 0.011 & 0.002 & 0.005 & 0.01 & 0.012 & \textbf{0.38} \\
        \textbf{0.5} & \textbf{0.412} & 0.369 & 0.329 & 0.395 & 0.393 & 0.348 & 0.351 \\
        \textbf{1.0} & 0.397 & \textbf{0.559} & 0.385 & 0.431 & 0.401 & 0.412 & 0.409 \\
        \hline \hline
        \multicolumn{8}{c}{Generator$_3$} \\
        \hline
        \textbf{0.0} &  & 0.324 & 0.38 & 0.381 & \textbf{0.395} & 0.39 & 0.38 \\
        \textbf{0.5} & \textbf{0.412} & 0.402 & 0.396 & 0.394 & \textbf{0.412} & 0.391 & 0.389 \\
        \textbf{1.0} & 0.397 & \textbf{0.417} & 0.415 & 0.413 & 0.407 & 0.409 & 0.407 \\
    \end{tabular}
    \caption{DKT \textbf{MCC} per data combination on the \textbf{SLP} dataset. 
    Rows stand for real data ratio, columns for synthetic data ratio. 
    Best values per real data ratio are highlighted.}
    \label{table:mcc_SLP}
\end{table}

Generator$_2$ shows mostly similar results for both datasets as Generator$_1$, with the one exception.
Here, we can observe that using a synthetic data ratio of 3 without real data achieves results comparable to using real data only, see Figure \ref{fig:oulad_mae} and \ref{fig:SLP_mae}.
This observation is also confirmed for ACC and MCC in Table  \ref{table:acc_SLP} and \ref{table:mcc_SLP}.

For Generator$_3$, it can be seen that adding a smaller amount of synthetic data to the real data tends to slightly improve the MAE on OULAD and on SLP, see Figure \ref{fig:oulad_mae} and \ref{fig:SLP_mae}, as well as MCC and ACC as can be seen in tables \ref{table:acc_oulad}-\ref{table:mcc_SLP}.
But, adding more than 25-50\% of the real training set size worsens the model on the OULAD data set.
Using the synthetic data only shows to cause a test performance on par with using a lower amount of real data for both datasets.

\begin{table}[!th]
    \centering
    \begin{tabular}{l|ccc}
        & \textbf{MAE} & \textbf{MCC} & \textbf{ACC}\\
        \hline
        \textbf{>0.1} & 0.09 & 0 & 0.91\\
        \textbf{0.1} & 0.09 & 0 & 0.91\\
        \textbf{0.01} & 0.159 & -0.006 & 0.841\\
    \end{tabular}
    \caption{BKT results on the \textbf{OULAD} dataset. 
    Rows stand for real data ratio.
    Synthetic data ratio is 0 for all entries in the column.}
    \label{table:mcc_oulad_bkt}
\end{table}

\begin{table}[!th]
    \centering
    \begin{tabular}{l||c|cccccc}
        & \textbf{0} & \textbf{0.25} & \textbf{0.5} & \textbf{0.75} & \textbf{1} & \textbf{2} & \textbf{3}\\
        \hline\hline
        \multicolumn{8}{c}{Generator$_1$} \\
        \hline
        \textbf{0.0} &  & 0.103 & 0.103 & 0.103 & 0.103 & 0.103 & 0.103\\
        \textbf{0.5} & \textbf{0.896} & 0.889 & 0.862 & 0.805 & 0.739 & 0.24 & 0.106\\
        \textbf{1.0} & \textbf{0.897} & 0.895 & 0.89 & 0.881 & 0.863 & 0.733 & 0.51\\
        \hline \hline
        \multicolumn{8}{c}{Generator$_2$} \\
        \hline
        \textbf{0.0} &  & 0.897 & 0.897 & 0.897 & 0.897 & 0.897 & 0.897\\
        \textbf{0.5} & 0.896 & 0.896 & 0.896 & \textbf{0.897} & \textbf{0.897} & \textbf{0.897} & \textbf{0.897}\\
        \textbf{1.0} & 0.897 & 0.987 & 0.897 & 0.897 & 0.897 & 0.897 & 0.897\\
        \hline \hline
        \multicolumn{8}{c}{Generator$_3$} \\
        \hline
        \textbf{0.0} &  & 0.896 & 0.896 & 0.896 & 0.896 & \textbf{0.897} & \textbf{0.897}\\
        \textbf{0.5} & 0.896 & 0.896 & 0.896 & 0.896 & 0.896 & \textbf{0.897} & \textbf{0.897}\\
        \textbf{1.0} & 0.897 & 0.897 & 0.897 & 0.897 & 0.897 & 0.897 & 0.897\\
    \end{tabular}
    \caption{BKT \textbf{ACC} per data combination on the \textbf{SLP} dataset. 
    Rows stand for real data ratio, columns for synthetic data ratio. 
    Best values per real data ratio are highlighted.}
    \label{table:acc_SLP_BKT}
\end{table}

\begin{table}[!th]
    \centering
    \begin{tabular}{l||c|cccccc}
        & \textbf{0} & \textbf{0.25} & \textbf{0.5} & \textbf{0.75} & \textbf{1} & \textbf{2} & \textbf{3}\\
        \hline\hline
        \multicolumn{8}{c}{Generator$_1$} \\
        \hline
        \textbf{0.0} &  & 0.0 & 0.0 & 0.0 & 0.0 & 0.0 & 0.0\\
        \textbf{0.5} & -0.001 & -0.001 & \textbf{0.009} & 0.005 & 0.005 & -0.002 & -0.001\\
        \textbf{1.0} & -0.002 & -0.005 & 0.003 & 0.005 & \textbf{0.008} & 0.006 & 0.01\\
        \hline \hline
        \multicolumn{8}{c}{Generator$_2$} \\
        \hline
        \textbf{0.0} &  & 0.0 & 0.0 & 0.0 & 0.0 & 0.0 & 0.0\\
        \textbf{0.5} & -0.002 & \textbf{0.0} & \textbf{0.0} & -0.001 & -0.001 & -0.003 & 0.0\\
        \textbf{1.0} & -0.002 & -0.002 & -0.002 & -0.002 & -0.001 & \textbf{0.0} & \textbf{0.0}\\
        \hline \hline
        \multicolumn{8}{c}{Generator$_3$} \\
        \hline
        \textbf{0.0} &  & 0.0 & \textbf{0.002} & 0.0 & -0.002 & -0.002 & 0.0\\
        \textbf{0.5} & -0.002 & 0.0 & 0.0 & 0.0 & -0.003 & -0.002 & \textbf{0.001}\\
        \textbf{1.0} & \textbf{-0.002} & \textbf{-0.002} & \textbf{-0.002} & -0.003 & -0.003 & \textbf{-0.002} & -0.003\\
    \end{tabular}
    \caption{BKT \textbf{MCC} per data combination on the \textbf{SLP} dataset. 
    Rows stand for real data ratio, columns for synthetic data ratio. 
    Best values per real data ratio are highlighted.}
    \label{table:mcc_SLP_BKT}
\end{table}

The BKT on the OULAD dataset appears completely uninfluenced by the addition or substraction of data on the above scales.
Here each variation including the baseline achieves an MAE of 0.09, an ACC of 0.91, and an MCC of 0.
To test, if smaller fractions of real data make a difference, we also include a run on 10\% and 1\% of the real dataset and the results in Table \ref{table:mcc_oulad_bkt} show that only a small ratio of 1\% causes a change in the results.

On the SLP dataset, an effect can be seen using Generator$_1$, see Figure \ref{fig:SLP_mae_BKT}, as well as Table \ref{table:acc_SLP_BKT} and \ref{table:mcc_SLP_BKT}.
These show that the performance on synthetic data does not come close to that on real data and adding the synthetic data to the real one consistently worsens.
For the other Generators however, the results look much more like for the OULAD dataset, as different data ratios show essentially no effect.

\section{Discussion} \label{sec:disc}

Regarding the general grade distribution, all three Generators come close to the real data, as can be seen in figures \ref{fig:stats_fig3} - \ref{fig:stats_fig6}.
However, using the generated data for training KT models has little to no positive effect on the performance.
For Generator$_1$ this likely lies in the fact that the synthetic grades are sampled independently of each other and of any other context, therefore not being able to represent a student actually going through a number of exercises in sequence.
Generator$_2$ does include a dependence on the time step, at which each exercise is taken.
However, it still does not account for the dependence of grades in a sequence on each other.
Generator$_3$ eventually samples full sequences and adapting those, leaving some degree of dependency of the sampled grades intact.
This is reflected in the fact, that the performance of the DKT model trained on synthetic data only, is on par with the baseline performance.
Generator$_3$ is also the only one for which we can observe a slight improvement when adding synthetic data to the real one.

Another reason that the performance does not improve with the generated data might lie in the skew of the datasets.
Especially the OULAD dataset shows a bias towards good grades, as can be seen in for example in Figure \ref{fig:stats_fig3}.
And while this is not the case for the SLP dataset, the same figure does also show that the chosen distribution for this dataset does also result in a skew towards positive grades.
So adding this kind of data may cause the KT models to be more biased towards positive grades, not being able to predict negative grades in a more balanced test set.

The BKT does not appear to profit from the synthetic data at all.
Here, even the baselines for both datasets already show an MCC close to 0, so the model seems already more biased towards the majority class.
That may be a reason   why adding synthetic data to the model does not change anything in the OULAD dataset.

\section{Conclusion} \label{sec:conclusion}

Our Synthetic Data Generators have been shown to produce data with overall distributions close to the real data.
However just drawing from distributions does not provide enough additional information for the KT models to be able to improve performance using this data.
But we could produce synthetic data close enough to the real data to achieve KT performances close to the real data.

This gives hope that we can build on this work to achieve better synthetic datasets in future work.
Next steps could include adapting the sampling approach from Generator$_3$ to not be too close to sampling real data sequences.
Another step to build on this study might be to utilize the synthetic data from more directed sampling to explicitly balance out biased datasets similar to an oversampling approach.

\ack This work is part of the project THISuccess$^{AI}$ under
project number FBM2020-EA-1690-07540 and has been funded by the Stiftung Innovation in der Hochschullehre.
\newpage
\bibliography{ecai}

\begin{thebibliography}{10}

\bibitem{badrinath2021pybkt}
Anirudhan Badrinath, Frederic Wang, and Zachary Pardos, `pybkt: An accessible python library of bayesian knowledge tracing models', in {\em Proceedings of the 14th International Conference on Educational Data Mining}, pp. 468--474, (2021).

\bibitem{Chang_JunYi_2015}
Haw{-}Shiuan Chang, Hwai{-}Jung Hsu, and Kuan{-}Ta Chen, `Modeling exercise relationships in e-learning: A unified approach', in {\em International Conference on Educational Data Mining (EDM)}, (2015).

\bibitem{choi_ednet_2020}
Youngduck Choi, Youngnam Lee, Dongmin Shin, Junghyun Cho, Seoyon Park, Seewoo Lee, Jineon Baek, Chan Bae, Byungsoo Kim, and Jaewe Heo.
\newblock Ednet: A large-scale hierarchical dataset in education, 2020.

\bibitem{corbett1994knowledge}
Albert~T Corbett and John~R Anderson, `Knowledge tracing: Modeling the acquisition of procedural knowledge', {\em User modeling and user-adapted interaction}, {\bf 4},  253--278, (1994).

\bibitem{dai2021knowledge}
Miao Dai, Jui-Long Hung, Xu~Du, Hengtao Tang, and Hao Li, `Knowledge tracing: A review of available techniques.', {\em Journal of Educational Technology Development \& Exchange}, {\bf 14}(2), (2021).

\bibitem{efron1983estimating}
Bradley Efron, `Estimating the error rate of a prediction rule: improvement on cross-validation', {\em Journal of the American statistical association}, {\bf 78}(382),  316--331, (1983).

\bibitem{efron1997improvements}
Bradley Efron and Robert Tibshirani, `Improvements on cross-validation: the 632+ bootstrap method', {\em Journal of the American Statistical Association}, {\bf 92}(438),  548--560, (1997).

\bibitem{GDPR2016a}
{European Parliament} and {Council of the European Union}.
\newblock Regulation ({EU}) 2016/679 of the {European} {Parliament} and of the {Council}.

\bibitem{feng_assistment_2009}
Mingyu Feng, Neil Heffernan, and Kenneth Koedinger, `Addressing the assessment challenge with an online system that tutors as it assesses', {\em User Modeling and User-Adapted Interaction}, {\bf 19}(3),  243--266, (August 2009).

\bibitem{ghosh2020context}
Aritra Ghosh, Neil Heffernan, and Andrew~S Lan, `Context-aware attentive knowledge tracing', in {\em Proceedings of the 26th ACM SIGKDD international conference on knowledge discovery \& data mining}, pp. 2330--2339, (2020).

\bibitem{8924064}
Jianhua Han, Wei Zhao, Qiang Jiang, Mohamed Oubibi, and Xiangen Hu, `Intelligent tutoring system trends 2006-2018: A literature review', in {\em 2019 Eighth International Conference on Educational Innovation through Technology (EITT)}, pp. 153--159, (2019).

\bibitem{kaser2017dynamic}
Tanja K{\"a}ser, Severin Klingler, Alexander~G Schwing, and Markus Gross, `Dynamic bayesian networks for student modeling', {\em IEEE Transactions on Learning Technologies}, {\bf 10}(4),  450--462, (2017).

\bibitem{kuzilek_2017_oulad}
Jakub Kuzilek, Martin Hlosta, and Zdenek Zdrahal, `Open {University} {Learning} {Analytics} dataset', {\em Scientific Data}, {\bf 4}(1),  170171, (November 2017).

\bibitem{lee_NPAKT_2019}
Youngnam Lee, Youngduck Choi, Junghyun Cho, Alexander~R. Fabbri, Hyunbin Loh, Chanyou Hwang, Yongku Lee, Sang-Wook Kim, and Dragomir Radev.
\newblock Creating {A} {Neural} {Pedagogical} {Agent} by {Jointly} {Learning} to {Review} and {Assess}, 2019.

\bibitem{liu2019ekt}
Qi~Liu, Zhenya Huang, Yu~Yin, Enhong Chen, Hui Xiong, Yu~Su, and Guoping Hu, `Ekt: Exercise-aware knowledge tracing for student performance prediction', {\em IEEE Transactions on Knowledge and Data Engineering}, {\bf 33}(1),  100--115, (2019).

\bibitem{liu2021survey}
Qi~Liu, Shuanghong Shen, Zhenya Huang, Enhong Chen, and Yonghe Zheng, `A survey of knowledge tracing', {\em arXiv preprint arXiv:2105.15106}, (2021).

\bibitem{liu_QEmbKT_2020}
Yunfei Liu, Yang Yang, Xianyu Chen, Jian Shen, Haifeng Zhang, and Yong Yu.
\newblock Improving {Knowledge} {Tracing} via {Pre}-training {Question} {Embeddings}, 2020.

\bibitem{matthews_1975}
B.W. Matthews, `Comparison of the predicted and observed secondary structure of t4 phage lysozyme', {\em Biochimica et Biophysica Acta (BBA) - Protein Structure}, {\bf 405}(2),  442--451, (1975).

\bibitem{nakagawa2019graph}
Hiromi Nakagawa, Yusuke Iwasawa, and Yutaka Matsuo, `Graph-based knowledge tracing: modeling student proficiency using graph neural network', in {\em IEEE/WIC/ACM International Conference on Web Intelligence}, pp. 156--163, (2019).

\bibitem{pandey_SAKT_2019}
Shalini Pandey and George Karypis.
\newblock A {Self}-{Attentive} model for {Knowledge} {Tracing}, 2019.

\bibitem{pardos_assistment_2014}
Zach~A Pardos, Ryan~S.J.D Baker, Maria San~Pedro, Sujith~M Gowda, and Supreeth~M Gowda, `Affective {States} and {State} {Tests}: {Investigating} {How} {Affect} and {Engagement} during the {School} {Year} {Predict} {End}-of-{Year} {Learning} {Outcomes}', {\em Journal of Learning Analytics}, {\bf 1}(1),  107--128, (May 2014).

\bibitem{piech_DKT_2015}
Chris Piech, Jonathan Bassen, Jonathan Huang, Surya Ganguli, Mehran Sahami, Leonidas~J Guibas, and Jascha Sohl-Dickstein, `Deep {Knowledge} {Tracing}', in {\em Advances in {Neural} {Information} {Processing} {Systems}}, eds., C.~Cortes, N.~Lawrence, D.~Lee, M.~Sugiyama, and R.~Garnett, volume~28. Curran Associates, Inc., (2015).

\bibitem{taskesen_erdogan_2023_7767074}
Erdogan Taskesen.
\newblock {distfit is a python library for probability density fitting.}, March 2023.
\newblock {If you use this software, please cite it using these metadata.}

\bibitem{yu_2021_slp}
Ziding Shen Penghe Chen Xiaoqing Li Shengquan~Yu Yu~Lu, Yang~Pian, `{SLP}: {A} {Multi}-{Dimensional} and {Consecutive} {Dataset} from {K}-12 {Education}', in {\em 29th {International} {Conference} on {Computers} in {Education} ({ICCE})}, (2021).

\bibitem{yudelson2013individualized}
Michael~V Yudelson, Kenneth~R Koedinger, and Geoffrey~J Gordon, `Individualized bayesian knowledge tracing models', in {\em Artificial Intelligence in Education: 16th International Conference, AIED 2013, Memphis, TN, USA, July 9-13, 2013. Proceedings 16}, pp. 171--180. Springer, (2013).

\end{thebibliography}
\end{document}